\documentclass[aps,pra,twocolumn,showpacs,amsmath,amssymb,groupedaddress]{revtex4-1}
\usepackage{times}
\usepackage{graphicx}
\begin{document}
\title{Anisotropic Superfludity in the Two-Species Polar Fermi Gas}
\author{Renyuan Liao}
\author{Joachim Brand}
\affiliation{Institute for Advanced Study and Centre for Theoretical Chemistry and Physics, Massey University, Auckland 0632, New Zealand}
\date{\today}

\begin{abstract}
We study the superfluid pairing in a two-species gas of  heteronuclear fermionic molecules with equal density. The interplay of the isotropic $s$-wave interaction and anisotropic long-range dipolar interaction reveals rich physics.
We find that the single-particle momentum distribution has a characteristic ellipsoidal shape that can be reasonably represented by a deformation parameter $\alpha$ defined similarly to the normal phase. Interesting momentum-dependent features of the order parameter are identified. We calculate the critical temperatures of both the singlet and triplet superfluid, suggesting a possible pairing symmetry transition by tuning the s-wave or dipolar interaction strength.
\end{abstract}
\maketitle
\section{Introduction}
The recent experimental realization and coherent control of high phase-space density quantum gas  of the polar molecules $^{40}K^{87}Rb$~\cite{JIN08,YE08,JIN10}  provides an excellent opportunity to study the effects of anisotropic long-range dipole-dipole interactions. Theoretical proposals employing degenerate polar molecules range from the study of exotic quantum phases of matter~\cite{FRAD09,SAM06} and quantum gas dynamics~\cite{TL09,SL00} to quantum simulations of highly correlated condensed matter systems~\cite{PUP09} and schemes for quantum information processing~\cite{AND06}.

Two fundamental properties of dipolar Fermi gases are superfluid pairing~\cite{BARA02,BARA04,CON10} and Fermi surface deformation~\cite{MIY08,SOG09,ZHA09}, originating from the partially attractive nature of the dipolar interaction and anisotropic Fock exchange interaction. For dipolar Fermi gases with two hyperfine states, one can tune not only the dipole-dipole interaction by a fast rotating orienting field~\cite{STE02}, but also the s-wave inter-species interaction via a Feshbach resonance. Therefore, one expects that rich physics will emerge as a result of the interplay of the anisotropic long-range dipole interaction and short-range s-wave interaction.

In this work, we study BCS pairing by taking account of the Fock exchange term in a self-consistent way. We find that the anisotropic nature of the dipolar interaction leads to an anisotropic momentum space distribution of the number density and an anisotropic order parameter. We generalize the definition of the deformation parameter introduced in Ref.~\cite{MIY08} to describe the anisotropic number distribution in the pairing phase and find that it gives a good description. Interesting features of the order parameter in momentum space are revealed, manifesting fascinating consequences of the dipolar interaction. Competing effects of the contact s-wave interaction and the dipolar interaction are identified in the study of the transition temperature of the superfluid state, suggesting the possibility of tuning the pairing symmetry by tuning the dipolar interaction.

\section{Model}
We consider a homogeneous gas of two species  of fermionic heteronuclear molecules, $\sigma=\uparrow$ and $\downarrow$. For simplicity, we further assume that each species has the same mass and density. The electric dipoles of the molecules with moment $\mathbf{d}$  are oriented along the $z-$axis by a sufficiently strong external electric field such that the spin-independent part of the electronic dipole-dipole interaction becomes $V_{dd}(\mathbf{q})=(4\pi/3)d^2(3\cos^2{\theta_\mathbf{q}}-1)$, with $\theta_\mathbf{q}$ being the angle between momentum $\mathbf{q}$ and the direction of the $z$ axis in which the dipoles are aligned. In addition, we assume that molecules also interact via a contact interaction with strength $g$. This system is described by the following Hamiltonian
\begin{eqnarray}
  H-\mu n=\sum_{\mathbf{k}\sigma}(\epsilon_{k}-\mu)c_{\mathbf{k}\sigma}^\dagger c_{\mathbf{k}\sigma}+\nonumber\\
  \frac{1}{2\mathcal{V}}\sum_{\mathbf{kpq}\sigma \sigma^\prime}V_{\sigma \sigma^\prime}(\mathbf{q})c_{\mathbf{k}+\mathbf{q} \sigma}^\dagger c_{\mathbf{p}-\mathbf{q} \sigma^\prime}^\dagger c_{\mathbf{p} \sigma^\prime}c_{\mathbf{k}\sigma},
\end{eqnarray}
 where $\mu$ is the chemical potential, $n$ is the total number density, $\mathcal{V}$ is the volume, and $\epsilon_k=k^2/2m$ (where we have set $\hbar=1$). The interaction potential $V_{\sigma \sigma^\prime}(\mathbf{q})=g \delta_{\sigma, -\sigma^\prime} + V_{dd}(\mathbf{q})$ contains both dipole-dipole and contact interactions. Anticipating the importance of the Fock exchange term, we decouple the interaction in all three channels~\cite{SIM06}: direct channel, exchange channel and Cooper channel, resulting in the following effective mean-field Hamiltonian
\begin{eqnarray}
    & &H_{MF}=\sum_{\mathbf{k}\sigma}\xi_{\mathbf{k}\sigma}c_{\mathbf{k}\sigma}^\dagger c_{\mathbf{k}\sigma}+\nonumber\\
    & &\frac{1}{2}\sum_{\mathbf{k}\sigma\sigma\prime}\left[\Delta_{\sigma\prime\sigma}^*(\mathbf{k})c_{-\mathbf{k}\sigma\prime}c_{\mathbf{k}\sigma} +\Delta_{\sigma\prime\sigma}(\mathbf{k})c_{\mathbf{k}\sigma}^\dagger c_{-\mathbf{k}\sigma\prime}^\dagger\right].\label{eq:h}
 \end{eqnarray}
 Here $\xi_{\mathbf{k}\sigma}=\epsilon_\mathbf{k}-\mu+gn/2+\Sigma_{k\sigma}$, with self-consistent mean-fields defined as
\begin{eqnarray}
      \Sigma_{\mathbf{k}\sigma}=-\frac{1}{\mathcal{V}}\sum_\mathbf{p} V_{dd}(\mathbf{p}-\mathbf{k})<c_{\mathbf{p}\sigma}^\dagger c_{\mathbf{p}\sigma}>,\\
      \Delta_{\sigma\prime\sigma}(\mathbf{k})=\frac{1}{\mathcal{V}}\sum_\mathbf{p}
    V_{\sigma \sigma^\prime}(\mathbf{k}-\mathbf{p})<c_{-\mathbf{k}\sigma\prime}c_{\mathbf{k}\sigma}>.
 \end{eqnarray}
 Some comments are in order: The contact interaction affects the single-particle spectrum by shifting the chemical potential, which may be redefined as $\tilde{\mu}=\mu-gn/2$. The self energy $\Sigma_{k\sigma}$ encodes the anisotropic dipolar contribution from the Fock exchange term to the dressed single particle spectrum, which justifies our treatment. In addition, both parts of the interaction contribute to the pairing field.

The Hamiltonian~(\ref{eq:h}) is diagonalized by invoking the Bogoliubov transformation~\cite{VOL90}. We obtain self-consistent equations for the self energy, the order parameters and the number density
 \begin{eqnarray}
 \Sigma_{\mathbf{k}\sigma}=-\sum_\mathbf{p} V_{dd}(\mathbf{p}-\mathbf{k})\left[\frac{1}{2}-\frac{\xi_{\mathbf{k}\sigma}}{2E_{\mathbf{k}\sigma}}\tanh{\frac{\beta E_{\mathbf{k}\sigma}}{2}}\right],\label{eq:se}\\
 \Delta_{\sigma\sigma^\prime}(\mathbf{k})=-\sum_\mathbf{p}V_{\sigma \sigma^\prime}(\mathbf{k-p})\frac{\Delta_{\sigma\sigma^\prime}(\mathbf{p})}{2E_{\mathbf{k}\sigma}}\tanh{\frac{\beta E_{\mathbf{k}\sigma}}{2}},\label{eq:gap}\\
 n=\sum_\mathbf{k}n_{\mathbf{k}}=\sum_\mathbf{k\sigma}\frac{1}{2}\left[1-\frac{\xi_{\mathbf{k}\sigma}}{E_{\mathbf{k}\sigma}}\tanh{\frac{\beta E_{\mathbf{k}\sigma}}{2}}\right]\label{eq:number},
 \end{eqnarray}
 where $E_{\mathbf{k}\sigma}=\sqrt{\xi_{k\sigma}^2+\sum_{\sigma^\prime}|\Delta_{\sigma\sigma^\prime}(\mathbf{k})|^2}$ is the quasi-particle spectrum and $\beta=1/k_BT$ is the inverse temperature. Equation~(\ref{eq:gap}) formally diverges. For the contact interaction, one can eliminate the interaction strength $g$ in terms of the $s$-wave scattering length $a_s$ using $1/g=m/(4\pi a_s)-\frac{1}{\mathcal{V}}\sum_\mathbf{k}\frac{1}{2\epsilon_\mathbf{k}}$. The dipolar interaction can be regularized by replacing the bare interaction $V_{\sigma \sigma^\prime}(\mathbf{k-p})$ with the vertex function $\Gamma_{\sigma\sigma^\prime}(\mathbf{k-p})$ as explained in Refs.~\cite{BARA02} and ~\cite{GUR07}. To first order in the Born approximation, the gap equations become
\begin{equation*}\label{}
\Delta_{\sigma\sigma^\prime}(\mathbf{k})=\sum_\mathbf{p}V_{\sigma \sigma^\prime}(\mathbf{k-p})\left[\frac{\tanh{\beta E_{\mathbf{k}\sigma}}}{2E_{\mathbf{k}\sigma}}-\frac{1}{2\epsilon_\mathbf{k}}\right]\Delta_{\sigma\sigma^\prime}(\mathbf{p}).
\end{equation*}
The above equation, together with Eqs.~(\ref{eq:se}) and~(\ref{eq:number}) comprise a complete description of the dipolar Fermi gas and needs to be solved self-consistently.

 Due to the symmetry of the interaction potential, the momentum distribution, order parameter and self energy possess azimuthal symmetry in thermal equilibrium, as can be seen from the self-consistent equations by integrating out the azimuthal degree of freedom $\phi_\mathbf{k}$. Thus, the physical quantities $n_\mathbf{k}$, $\Delta_{\sigma\sigma^\prime}(\mathbf{k})$ and $\Sigma_{\mathbf{k}\sigma}$ are only functions of $(k,\theta_\mathbf{k})$. Numerically, we parameterize these quantities by two-dimensional grids living on the  domain $\Omega=[0,k_c]\times[0,\pi]$, where $k_c$ is the momentum cutoff. For spin singlet pairing the order parameter possesses inversion symmetry as well as azimuthal symmetry, $\Delta(k,\theta_\mathbf{k})=\Delta(k,\pi-\theta_\mathbf{k})$, while for spin triplet pairing $\Delta(k,\theta_\mathbf{k})=-\Delta(k,\pi-\theta_\mathbf{k})$. In our calculation, we parameterize the dipolar interaction by the dimensionless coupling parameter $C_{dd}=md^2(n_\sigma)^{1/3}$, where $n_\sigma=k_F^3/(6\pi^2)=n/2$, and the Fermi energy is $E_F=k_F^2/2m$. In addition to the azimuthally symmetric solution, vortex states with azimuthally varying phase ($\arg \Delta = i \nu \varphi$, where $\varphi$ is the azimuthal angle and $\nu$ is integer) are expected to exist and can be treated by straightforward generalization of the presented approach.

\section{Spin singlet pairing}
\begin{figure}[t]
{\scalebox{0.30}{\includegraphics[clip,angle=0]{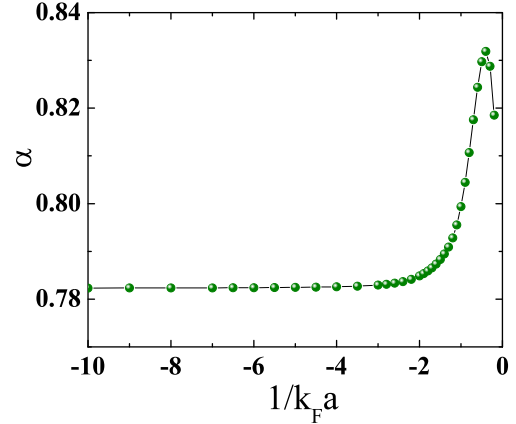}}}
\caption{(Color online) Deformation parameter $\alpha$ as a function of $s$-wave coupling strength $1/k_Fa$ at $C_{dd}=1$ for the singlet superfluid. }
\label{Fig1}
\end{figure}

 To investigate the interplay of the contact  and dipolar interactions, we devote in this paragraph to studying spin singlet pairing at zero temperature. In the normal phase, the Fermi surface of the dipolar gas has an ellipsoid shape~\cite{MIY08}: $n(\mathbf{k})=\Theta(k_F^2-\alpha^2k_z^2-k_x^2/\alpha-k_y^2/\alpha)$, where $\alpha$ is the deformation parameter. In the superfluid phase, we propose to use the deformation parameter $\alpha$ to measure the anisotropy of the  single-particle momentum distribution with a similar strategy. The angular distribution $n_\theta(\theta_\mathbf{k})$ can be obtained by integrating out the magnitude and azimuthal angle of the momentum $n_\theta(\theta_\mathbf{k})=\int n(\mathbf{k})k^2dk/(2\pi^2)$.
Since $n_\theta(0)/n_\theta(\pi/2)=k_z^3/k_x^3$ and $\alpha^2k_z^2=k_x^2/\alpha$, we can deduce that $\alpha=\left[n_\theta(\pi/2)/n_\theta(0)\right]^{2/9}$. In Fig.~\ref{Fig1} the deformation parameter $\alpha$ is plotted as a function of the $s$-wave coupling strength, characterized by $1/k_Fa_s$, at fixed dipole interaction strength $C_{dd}=1$. As the $s$-wave coupling strength is increased, the deformation parameter $\alpha$ increases before it reaches a maximum, signaling that the anisotropy of the single particle momentum distribution
is mitigated. It is energetically favorable for the system to become less anisotropic to benefit the energy gain from BCS pairing. The appearance of a pronounced peak suggests that there is an optimal value of the $s$-wave interaction strength to mitigate anisotropy.

\begin{figure}[t]
{\scalebox{0.30}{\includegraphics[clip,angle=0]{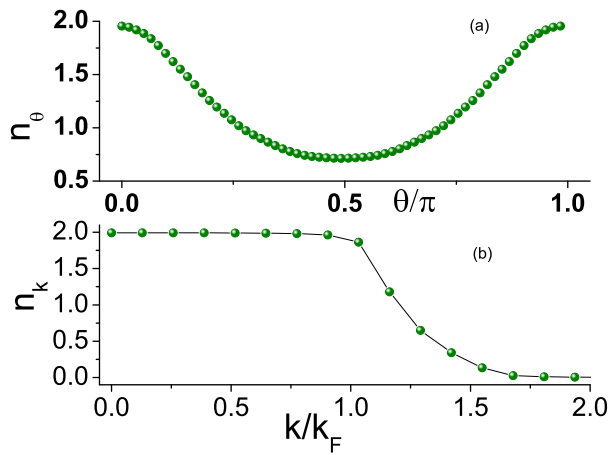}}}
\caption{(Color online) (a) Momentum integrated angular number distribution; (b) Angle-averaged momentum number distribution $n_k(k)=\int n(\mathbf{k})d\Omega/ 4\pi$ at $1/k_Fa=-1$ and $C_{dd}=1$ for the singlet superfluid.}
\label{Fig2}
\end{figure}

\begin{figure}[t]
{\scalebox{0.30}{\includegraphics[clip,angle=0]{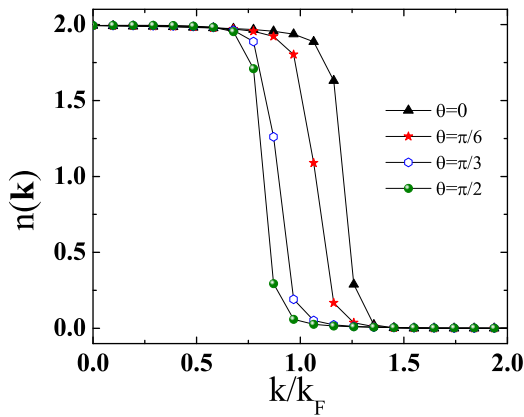}}}
\caption{(Color online) Radial number distribution in momentum space at $1/k_Fa=-1$ and $C_{dd}=1$ and different polar angles for the singlet superfluid.}
\label{Fig3}
\end{figure}

The single-particle momentum distribution is shown in Fig.~\ref{Fig2}. The angular distribution $n_\theta$ is symmetrical with respect to the orbital axis, with $n_\theta(\theta_\mathbf{k})=n_\theta(\pi-\theta_\mathbf{k})$ extending to the polar axis with a parabolic shape, as suggested in panel (a). The angle-averaged momentum  distribution $n_k(k)=\int n(\mathbf{k})d\Omega/4\pi$ is shown in panel (b). For $k<k_F$, the quantum states are essentially fully occupied. For $k>k_F$ the occupation number decreases rapidly with a $k^{-4}$ tail,
as can be derived from Eq.~(\ref{eq:number}). This $k^{-4}$ asymptote is valid for wave numbers smaller than the inverse range of the two-particle interaction. It is characteristic of BCS-type pairing and is also predicted in isotropic Fermi gases~\cite{GIO08,TAN08}.
To further reveal the anisotropic nature of the momentum distribution we plot the radial momentum distribution for different polar angles $\theta_\mathbf{k}=0, \pi/6, \pi/3, \pi/2$ in Fig.~\ref{Fig3}.
For any angle $\theta_\mathbf{k}$, the momentum distribution drops very sharply from almost full occupation to very small values at a critical value $k_c$.
At $\theta_\mathbf{k}=0$ the value of $k_c$ is the smallest. As one increases the polar angle from $0$ to $\pi/2$, the critical momentum increases monotonically. The critical momentum $k_c(\theta_\mathbf{k})$ assumes a similar role as the Fermi surface for the normal Fermi gas and defines an ellipsoidal shape with the long axis pointing along the $z$ axis and the short axis in the $x-y$ plane. Thus, our definition of the deformation parameter $\alpha$ above is justified.

\begin{figure}[t]
{\scalebox{0.30}{\includegraphics[clip,angle=0]{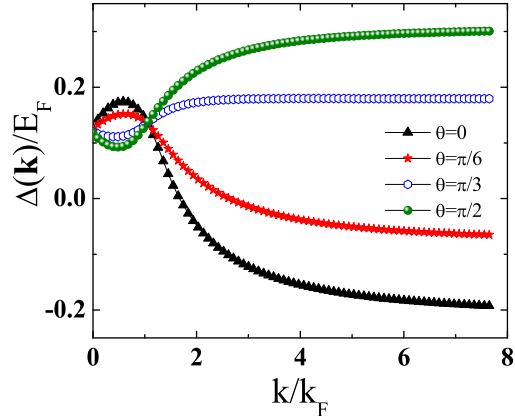}}}
\caption{(Color online) Radial gap parameter profile in momentum space at $1/k_Fa=-1$ and $C_{dd}=1$ and different polar angles for the singlet superfluid.}
\label{Fig4}
\end{figure}

Let us turn the attention to the anisotropy of the order parameter, which reveals the excitation spectrum of the system. As shown in Fig.~\ref{Fig4}, the order parameter shows interesting features in the momentum space: the order parameter is a non-monotonic function of the magnitude of momentum along fixed polar angle, a feature similar to the one found~\cite{BARA04} in the spatial distribution of order parameter for trapped polar molecules. For different polar angles the magnitude of order parameter coincides at two typical momentum magnitudes, approximately $0$ and $k_F$. Around the Fermi surface ($k\approx k_F$) the slope ${\rm d}\Delta/{\rm d}k$ increases monotonically from negative values for $\theta_\mathbf{k}=0$ to positive values for $\theta_\mathbf{k}=\pi/2$.
The isotropy of the order parameter at the Fermi surface and at zero momentum is a striking feature of the solutions of the mean-field Hamiltonian (2).

\section{Critical temperature}
 At $T\geq T_c$, the gap parameter vanishes and the system undergoes a 2nd order phase transition to the normal state. For spin triplet pairing the order parameters $\Delta_{\sigma\sigma^\prime}(\mathbf{k})$ are decoupled, satisfying the gap equations with exactly the same form. For convenience, to determine the critical temperature, we denote the order parameter $\Delta_{\sigma\sigma^\prime}(\mathbf{k})$ by $\Delta_\mathbf{k}$, regardless of the pairing spin symmetry. As $T$ approaches $T_c$ from below, the gap parameter is a small parameter, and the gap equation can be linearized:
\begin{eqnarray*}
  \Delta_\mathbf{k}&=&\sum_\mathbf{p}V(\mathbf{p}-\mathbf{k})\left[1/2\epsilon_\mathbf{p}-\frac{\tanh(\beta|\xi_\mathbf{p}|/2)}{2|\xi_\mathbf{p}|}\right]\Delta_\mathbf{p}.
\end{eqnarray*}

We define
\begin{eqnarray*}
   \mathcal{R}(\mathbf{p},\mathbf{k})=V(\mathbf{p}-\mathbf{k})\left[1/2\epsilon_\mathbf{p}-\frac{\tanh(\beta|\xi_\mathbf{p}|/2)}{2|\xi_\mathbf{p}|}\right],
\end{eqnarray*}
then one gets an eigenvalue equation
\begin{equation}\label{eq:eig}
  \Delta_\mathbf{k}=\sum_\mathbf{p}\mathcal{R}(\mathbf{p},\mathbf{k})\Delta_\mathbf{p}.
\end{equation}
Let us define the orthonormal function $\phi_l(\cos{\theta})=\sqrt{\frac{2l+1}{2}}P_l(\cos{\theta})$ and expand the gap parameter in terms of this basis functions $\Delta_\mathbf{k}=\sum_l \Delta_l(k)\phi_l(\theta_\mathbf{k})$. We can rewrite the Eq.~(\ref{eq:eig}) as
\begin{eqnarray*}
  \Delta_l(k)=\frac{\int p^2dp}{(2\pi)^2}\sum_{l^\prime}\mathcal{R}_{kl,pl^\prime}\Delta_{l^\prime}(p),
\end{eqnarray*}
where
\begin{align}
\nonumber
    \mathcal{R}_{kl,pl^\prime}&= \\
& \hspace{-5mm}\int {\rm d}(\cos\theta_\mathbf{k})\int  {\rm d}(\cos\theta_\mathbf{p})\mathcal{R}(\mathbf{p}-\mathbf{k})\phi_l(\cos\theta_\mathbf{k})\phi_{\l^\prime}(\cos\theta_\mathbf{p}).
\end{align}
In matrix notation we may write
\begin{eqnarray}
    \tilde{\Delta}=\tilde{\mathcal{R}}\tilde{\Delta},\\
    \tilde{\mathcal{R}}_{kl,pl^\prime}=\frac{p^2\delta p}{(2\pi)^2}\mathcal{R}_{kl,pl^\prime},\\
    \tilde{\Delta}_{kl}=\Delta_l(k).
\end{eqnarray}
In this equation, the matrix and vector indices are jointly labeled by the momentum magnitude $k$ and the orbital index $l$. For singlet pairing $l$ runs over even and positive integers while for triplet pairing $l$ is odd. The critical temperature $T_c$ can be found from the solutions of the secular equation
\begin{eqnarray}
    \mathbf{Det}(\tilde{\mathcal{R}}-\mathbf{I})=0.
\end{eqnarray}
\begin{figure}[t]
{\scalebox{0.30}{\includegraphics[clip,angle=0]{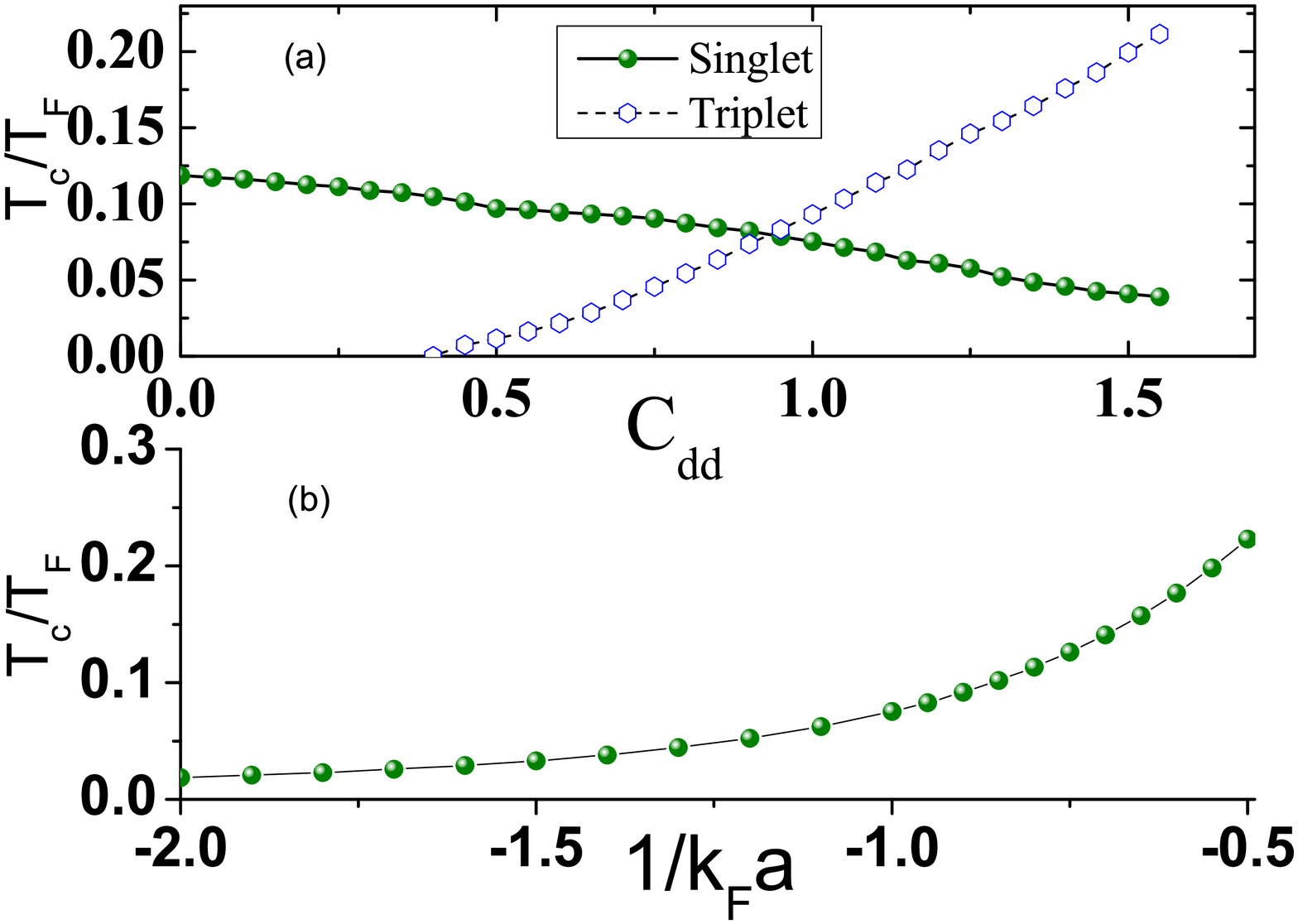}}}
\caption{(Color online) (a) Critical temperature as a function of dipole-dipole interaction strength $C_{dd}$ for singlet superfluid at $1/k_Fa=-1$ and for triplet superfluid; (b) critical temperature as a function of $1/k_Fa$ at $C_{dd}=1$ for singlet superfluid.}
\label{Fig5}
\end{figure}
We show the calculated critical temperature in Fig.~\ref{Fig5}. In panel (a), for fixed $s$-wave coupling with $1/k_Fa=-1$, the critical temperature $T_c$ of the singlet superfluid decreases monotonically as the dipole interaction strength increases. This is due to competing effects of the $s$-wave  and dipolar interactions. While the $s$-wave interaction favors a spherical Fermi surface over an ellipsoid Fermi surface favored by the dipolar interaction.  In the case of spin triplet pairing, $T_c$ increases with the dipolar interaction as expected since the $s$-wave interaction does not participate in the pairing mechanism. When the system is cooled down from the normal phase, the actual pairing symmetry, spin singlet or spin triplet, is determined by which of the critical temperatures are higher. By tuning the dipolar interaction strength $C_{dd}$, the quantum phase transition between singlet and triplet superfluids may be realized. In panel (b), for $C_{dd}=1$, the critical temperature for the singlet superfluid is clearly a monotonic function of the s-wave coupling strength. Here we only show $T_c$ in the range of weak interaction where the mean-field description gives quantitative reasonable results. For strong interaction at finite temperature, the fluctuation contribution is significant and one needs to resort to methods beyond the mean-field description.

\section{Experimental signature}
The anisotropic nature of the superfluid order parameter and momentum distribution bear consequences for experimental observations. The single-particle momentum distribution of trapped Fermi gases is routinely observed by time-of flight measurements~\cite{REG05}. Collecting angle-resolved data from a gas released from an axially symmetric trap should reveal the anisotropies predicted in Section III. Anisotropic features of the pairing gap and further details of the single-particle spectrum could be probed by radio frequency spectroscopy~\cite{CHI04,GAE10}.

\section*{Acknowledgement}This work was partially funded by the Marsden Fund (contract MAU0706) administered by the Royal Society of New Zealand.

\end{document}